\documentclass[pre,reprint,aps,raggedbottom]{revtex4-2}
\usepackage{amssymb}
\usepackage{physics}
\usepackage{float}
\usepackage{comment}
\usepackage{amsmath}
\usepackage{graphicx}
\usepackage{amsthm}

\usepackage{booktabs}
\usepackage{tikz}

\newtheorem{theorem}{Theorem}[section]

\newtheorem{example}[theorem]{Example}

\newcommand{\Des}{\operatorname{Des}}
\newcommand{\maj}{\operatorname{maj}}
\newcommand{\inv}{\operatorname{inv}}

\newcommand{\sgnrep}{\mathrm{sgn}}
\newcommand{\sgn}{\operatorname{sgn}}
\newcommand{\SYT}{\operatorname{SYT}}
\newcommand{\SSYT}{\operatorname{SSYT}}

\newcommand{\Var}{\operatorname{Var}}

\newcommand{\ZZ}{\mathbb{Z}}
\newcommand{\NN}{\mathbb{N}}

\usetikzlibrary{arrows.meta}
\definecolor{lvl}{RGB}{160,160,160}
\newcommand{\dpan}[9]{
\begin{scope}[xshift=#1,yshift=#2]
  \node[font=\small] at (1.5,2.62) {#3};
  \foreach \x/\lab in {0/1,1/2,2/3,3/4} {\node[font=\small] at (\x,2.08) {$i_{\lab}$};}
  \node[font=\small] at (0.5,2.08) {#4};
  \node[font=\small] at (1.5,2.08) {#5};
  \node[font=\small] at (2.5,2.08) {#6};
  \foreach \y in {0,0.4,0.8,1.2} {\draw[lvl,line width=0.6pt] (-0.12,\y) -- (3.12,\y);}
  \foreach \y/\lab in {0/1,0.4/2,0.8/3,1.2/4} {\node[font=\scriptsize,anchor=east] at (-0.26,\y) {\lab};}
  \node[font=\small] at (1.5,1.58) {#7};
  \node[font=\footnotesize] at (1.5,-0.62)
    {$\textstyle\sum i_k = #8,\quad \sum i_k^{2} = #9$};
\end{scope}}

\begin{document}
\title{The Arithmetic of Spectra: Factorization, Statistics, and Symmetric Functions}
\author{A. Chaudhary}
\email{aarifchaudharyg@gmail.com}
\affiliation{Hendrix Industries, Sealy, Texas 77474, USA}

\begin{abstract}
We study factorizations of the single-particle spectrum of non-interacting quantum systems and their consequences for many-particle statistics. Representing the single-particle partition function by a spectral alphabet, tensor factorizations become multiplicative factorizations of the alphabet, which can be lifted to canonical Bose and Fermi partition functions using standard symmetric-function and $\lambda$-ring identities. We show how product spectra arise from different factorizations, how antisymmetrization can be assigned across an odd number of factors, and which tensor factorizations are compatible with a fixed spectrum. The paper therefore studies spectral factorization and provides a combinatorial framework for canonical Bose and Fermi partition functions.

\end{abstract}

\maketitle

\section{Introduction}
\label{sec:intro}

For noninteracting identical particles the canonical partition functions at fixed particle number satisfy a recursion driven by the single-particle partition function evaluated at multiples of $\beta$, which identifies $Z_{n,B}$ and $Z_{n,F}$ with the complete homogeneous and elementary symmetric functions of one and the same set of single-particle Boltzmann weights. The correspondence was noted by Ford\cite{Ford71}, the recursion derived from the cycle decomposition of the permutation sum by Borrmann and Franke\cite{BorrmannFranke1993} and from the path integral by Brosens, Devreese and Lemmens\cite{Borsens1997}, and the symmetric-function identification developed by Schmidt and Schnack\cite{SchmidtSchnack2002} and, in group-character form, by Balantekin\cite{Balantekin}. It has since been developed into a framework for occupation-number statistics and level correlations at arbitrary
degeneracy\cite{Barghathi2020}, and at finite imaginary-time discretization\cite{newtonsidentity}.

In this work we study arithmetic operations on the single-particle spectrum and their consequences for many-body thermodynamics. A spectrum is represented by the formal alphabet $X$ whose Adams operations $\psi^n(X)$ reproduce the power traces entering Newton's identities. Arithmetic operations such as addition and multiplication of alphabets then encode direct-sum and tensor-product structures of the underlying Hilbert space. We show how product spectra arise from factorizations that need not be spatial and how the antisymmetrization distinguishing the two statistics may be assigned across any odd number of factors. Furthermore, since the canonical thermodynamics depends only on $X$ and not on how $X$ is realized as a product, we answer the question of which tensor factorizations are compatible with a fixed spectrum. Lifting these factorizations to the canonical partition functions uses only standard symmetric-function and $\lambda$-ring identities, and assembling them gives a systematic calculus for quantum-statistical spectra in which bosonic and fermionic statistics are the operations $\sigma^n(X)$ and $\lambda^n(X)$ on one object, related by $\sigma^n(X)=(-1)^n\lambda^n(-X)$.

The paper is organized as follows. Sec.\ref{sec:establish} sets up the spectral alphabet and the $\lambda$-ring dictionary for the canonical partition functions. Sec.\ref{sec:DirectSum} treats additive spectra. Sec.\ref{sec:twofactor} develops the two-factor case, where the Cauchy and dual Cauchy identities resolve the two statistics into symmetry sectors and the quasisymmetric refinement makes their interchange a local operation on occupation configurations, and Sec.\ref{sec:realizations} gives physical realizations beyond spatial separability. Sec.\ref{sec:kronecker} extends this to an arbitrary number of factors through generalized Kronecker coefficients, where the assignment of antisymmetrization across an odd number of factors appears. Sec.\ref{sec:tensor_resolution} asks which factorizations a fixed spectrum admits, and Sec.\ref{sec:operator-extension} generalizes the construction from the imaginary-time propagator to an arbitrary trace-class one-particle kernel. Appendix~\ref{sec:highharmonic} specializes the calculus to the harmonic trap, where geometric factors collapse the general expansions into exchange polynomials whose reflection gives the Bose-Fermi temperature relation in odd dimensions but leaves the fermion sign problem intact.

\section{Spectra as Alphabets}
\label{sec:establish}
For $n$ identical noninteracting particles with single-particle partition function $Z_1(\beta)$, the canonical partition functions obey
\begin{align}
 Z_n^{B}&=\frac1n\sum_{k=1}^{n}z_k\,Z_{n-k}^{B},
 \label{eq:newtonB}\\
 Z_n^{F}&=\frac1n\sum_{k=1}^{n}(-1)^{k-1}z_k\,Z_{n-k}^{F},
 \label{eq:newtonF}
\end{align}
with $z_k=Z_1(k\beta)$. For $d$-dimensional harmonic trap,
\begin{equation}
 (z_n)^d=\left(\frac{b^{n/2}}{1-b^n}\right)^{d},
 \qquad
 b=e^{-\beta},
 \label{eq:zharmonic}
\end{equation}
in units $\hbar\omega=1$. In Ref.\cite{ChaudharyValenzuelaChin2026} the same expressions were established for an arbitrary number of imaginary-time slices and for any choice of short-time propagator, with the discretization entering only through the value of $b$. Identifying the symmetric functions in \eqref{eq:newtonB} and \eqref{eq:newtonF} with $\lambda$-rings\cite{Yau2010} gives the dictionary
\begin{center}
\begin{tabular}{ccccc}
$\lambda$-ring && $Z_n$ && symmetric functions\\
\hline
$\lambda^n(X)$ &$\leftrightarrow$& $Z^{d}_{n,F}$ &$\leftrightarrow$& $e_n$\\
$\sigma^n(X)$ &$\leftrightarrow$& $Z^{d}_{n,B}$ &$\leftrightarrow$& $h_n$\\
$\psi^n(X)$&$\leftrightarrow$& $z^{d}_{n}$&$\leftrightarrow$& $p_n$\\
$\lambda_t(X)$ &$\leftrightarrow$& $\Xi^{d}_{F}(t)$ &$\leftrightarrow$& $E(t)$\\
$\sigma_t(X)$ &$\leftrightarrow$& $\Xi^{d}_{B}(t)$ &$\leftrightarrow$& $H(t)$\\
\end{tabular}
\end{center}
where $\Xi^{d}_{F}(t)=\sum_{k\ge0}Z^{d}_{k,F}t^{k}$ and likewise for $\Xi^{d}_{B}$. If $t$ is identified with the fugacity ($t=e^{\beta\mu}$), these are the fermionic and bosonic grand-canonical partition functions. The operation building them from the alphabet is the plethystic exponential. For
$X=\sum_a g_a b_a$,
\begin{equation}
\begin{split}
 \mathrm{PE}[X](t)&=\prod_a(1-t\,b_a)^{-g_a}
 \\&=\exp\Big[\sum_{r\ge1}\tfrac{t^r}{r}\psi^r(X)\Big]=\sigma_t(X),\\
 \mathrm{PE}_\Lambda[X](t)&=\prod_a(1+t\,b_a)^{g_a}
 \\&=\exp\Big[\sum_{r\ge1}\tfrac{(-1)^{r-1}t^r}{r}\psi^r(X)\Big]=\lambda_t(X),
\end{split}
\label{eq:PE}
\end{equation}
as formal series in $t$, the graded characters of the symmetric and exterior algebras on the one-particle space. The same operation organizes multi-trace operator counting in gauge theory\cite{Feng_2007}. Everything below is stated canonically through $Z^d_{n,B/F}=[t^n]\,\Xi^d_{B/F}$, but the grand canonical forms can be recovered at any point by resumming. 

The two families are not independent, and the relations can be seen through
\begin{equation}
\begin{split}
 \sigma_t(X)&=\lambda_{-t}(X)^{-1},
 \\
 \psi_{t}(X) &= \sum_{n\ge1}\psi^n(X)t^n=t\,\frac{d}{dt}\log\sigma_t(X),\\
 \psi_{-t}(X) &= -t \frac{d}{dt}\log\lambda_{t}(X).
 \label{eq:sigmadef}
\end{split}
\end{equation}
Therefore \eqref{eq:newtonB} and \eqref{eq:newtonF} are two ways of reading the same Adams operations $\psi^n$ off the two generating functions. In this notation \eqref{eq:newtonB} and \eqref{eq:newtonF} can be written as
\begin{equation}
\begin{split}
\sigma^n(x) &= \frac{1}{n}\sum_{k=1}^n \psi^k(x) \sigma^{(n-k)}(x), 
\\ \lambda^n(x) &= \frac{1}{n}\sum_{k=1}^n (-1)^{k-1}\psi^k(x) \lambda^{(n-k)}(x)
\end{split}
\end{equation}

For the harmonic oscillator, as shown in Ref.\cite{ChaudharyValenzuelaChin2026} this structure admits a concrete realization by taking $K=\mathbb{Q}[[b]]$, the ring of formal power series in $b$, equipped with its canonical $\lambda$-ring structure $\Lambda(K)=1+K[[t]]^{+}$. We factor the zero-point contribution $b^{nd/2}$ out of the recursion. Under the identification $\psi^n(X)\leftrightarrow z^{d}_n$,
\begin{equation}
z^{d}_n=\psi^n(X)=\sum_{k\ge0}\binom{k+d-1}{d-1}b^{nk},
 \label{eq:psiX}
\end{equation}
giving $\psi^n(b^\alpha)=b^{n\alpha}$ and $X=(1-b)^{-d}=z^{d}_1$. So $X$ is the single-particle Boltzmann alphabet of the trap. For a general non-interacting spectrum with levels $\epsilon_a$ of degeneracy $g_a$, let $b_a = e^{-\beta \epsilon_a}$, then the alphabet is the formal sum
\begin{equation}
 X=\sum_a g_a b_a\,,
 \qquad
 \text{and}\qquad \psi^n(X)=\sum_a g_a b^n_a\,.
 \label{eq:generalalphabet}
\end{equation}
The definition of the line elements (elements such that $\psi^n(x) = x^n$) being the Boltzmann weight $b$ is not arbitrary, but one that keeps the single particle spectrum intact. The bosonic and fermionic operations are related by a single identity. Since $\lambda_t(-X)=\lambda_t(X)^{-1}$, comparison with \eqref{eq:sigmadef} gives
\begin{equation}
\begin{split}
 \sigma^n(X)=&(-1)^n\lambda^n(-X) \\&\,\,\longrightarrow{}\,\, Z_{n,B}^d(X) = (-1)^n Z_{n,F}^d(-X)
 \label{eq:negation}
\end{split}
\end{equation}
i.e. the two statistics are the same operation evaluated at negated argument.

Although we focus on the imaginary-time propagator $e^{-\beta H_1}$, the algebraic construction depends only on the power traces of the underlying one-particle operator. Sec.\ref{sec:operator-extension} shows that more generally for any trace-class kernel $K$ the following analysis holds, and also shows in more detail the origin of the correspondence using permutation projections of the $n$ particle Hilbert Space $\mathcal H_n=\mathcal H_1^{\otimes n}$.

\section{Additive Spectra}
\label{sec:DirectSum}
We first consider the case where the one-particle Hilbert space decomposes as a direct sum rather than a tensor product. Let
\begin{equation}
\mathcal H_1=\bigoplus_{l}\mathcal H_l .
\end{equation}
At the level of spectral alphabets, this corresponds to the additive composition rule
\begin{equation}
X=\sum_{l}\hat X_l .
\end{equation}
Unlike the multiplicative case, where a product alphabet generates a nontrivial Schur decomposition, as we will show in the next section, the additive structure follows directly from the identities,
\begin{equation}
\begin{split}
\sigma_t[X+Y]
&=
\sigma_t[X]\sigma_t[Y],\\
\lambda_t[X+Y]
&=
\lambda_t[X]\lambda_t[Y].
\end{split}
\end{equation}
The $n$ particle partition then reads,
\begin{equation}
\begin{split}
\sigma^n[X+Y]
&=
\sum_{n_1+n_2=n}
\sigma^{n_1}[X]\sigma^{n_2}[Y], \\
\lambda^n[X+Y]
&=
\sum_{n_1+n_2=n}
\lambda^{n_1}[X]\lambda^{n_2}[Y].
\label{eq:sumEq}
\end{split}
\end{equation}
Thus, a direct-sum decomposition simply distributes the particles among the independent spectral sectors. No additional coupling between sectors is introduced, and the composition is completely captured by the product of the individual generating functions. Subtraction of alphabets is the same identity with a negated argument,
\begin{equation}
\begin{split}
\sigma^n[X-Y]&=\sum_{k=0}^n(-1)^k\lambda^k[Y]\,\sigma^{n-k}[X],\\
\lambda^n[X-Y]&=\sum_{k=0}^n(-1)^k\sigma^k[Y]\,\lambda^{n-k}[X],
\end{split}
\label{eq:subtraction}
\end{equation}
which are the relations between fermionic and bosonic auxiliary partition functions under level removal in Ref.\cite{Barghathi2020}. Both simply follow from \eqref{eq:negation}, so the interchange of the two statistics under level removal is the negation law rather than a separate phenomenon.

For a concrete realization of this, consider the case of Landau levels. Given that the potentials don't couple the levels, one can write the Hilbert space as a direct sum of the different levels $l \in \operatorname{LL}$,
\begin{equation*}
\mathcal H_1=\bigoplus_{l \in \operatorname{LL}}\mathcal H_l.
\end{equation*}
Hence, this gives the alphabet $X = \sum_l \hat X_l$ and the partition functions given by \eqref{eq:sumEq}.

\section{Two-Alphabet Product spectra}
\label{sec:twofactor}

We first focus on just studying the two-alphabet product spectra before presenting the general results. The two-alphabet spectra are much simpler to deal with since they don't require any Kronecker coefficients (we will introduce these in Sec.\ref{sec:kronecker}) and hence provide a better base to understand the physical implications of the framework while still providing relevant examples.

Let the single-particle spectrum with alphabet $X$ be separable into two factors, with alphabets
\begin{equation}
\begin{split}
 \hat X=&\sum_i g_{\hat x,i}\,\hat x_i,
 \qquad
 \hat Y=\sum_j g_{\hat y,j}\,\hat y_j,
 \\
 &\hat x_i=e^{-\beta\epsilon_i},
 \qquad
 \hat y_j=e^{-\beta\eta_j},
 \label{eq:twofactoralph}
\end{split}
\end{equation}
each a formal sum of letters with multiplicity in the sense of \eqref{eq:generalalphabet}. Separability means the single-particle energies are the sums $\epsilon_i+\eta_j$, so the full alphabet is the product
\begin{equation}
 X=\hat X\hat Y=\sum_{i,j}g_{\hat x,i}g_{\hat y,j}\,\hat x_i\hat y_j.
 \label{eq:twoalph}
\end{equation}
The product $g_{\hat x,i}g_{\hat y,j}$ may itself not be the degeneracy of the corresponding weight in $X$. Distinct pairs $(i,j)$ may carry the same total energy, therefore the degeneracy of a level of $X$ is obtained only after collecting them. Writing $x_k$ for the distinct values taken by the products $\hat x_i\hat y_j$,
\begin{equation}
 X=\sum_k g_k\,x_k,
 \qquad
 g_k=\sum_{(i,j)\,:\;\hat x_i\hat y_j=x_k}g_{\hat x,i}\,g_{\hat y,j}.
 \label{eq:productdegeneracy}
\end{equation}

While the identities of this section require only that the alphabet factor as $X=\hat X\hat Y$, reading the indices of a single factor as quantum numbers of a subsystem, as in the occupation language used later in this section and in Sec.\ref{sec:realizations}, requires in addition that the one-particle Hilbert space itself factors as $\mathcal H_1=\mathcal H_{\hat X}\otimes\mathcal H_{\hat Y}$ with the Hamiltonian $H_1=H_{\hat X}\otimes I+I\otimes H_{\hat Y}$.

The bosonic partition function of a product alphabet is resolved by the Cauchy identity \cite{Macdonald1995,StanleyEC2},
\begin{equation}
 \sigma^n(X)=\sum_{\lambda\vdash n}s_\lambda[\hat X]\,s_\lambda[\hat Y].
 \label{eq:Cauchy}
\end{equation}
The fermionic case can directly be obtained by negating the alphabet. In the Schur basis, the negation \eqref{eq:negation} reads \cite{LoehrRemmel2011}
\begin{equation}
 s_\lambda[-\hat X]=(-1)^{|\lambda|}s_{\lambda'}[\hat X],
 \label{eq:schurnegation}
\end{equation}
Negating the first factor in \eqref{eq:Cauchy} therefore gives
\begin{equation*}
\begin{split}
 &(-1)^n\lambda^n[\hat X\hat Y]=\sigma^n[-\hat X\hat Y] \\
 &=\sum_{\lambda\vdash n}s_\lambda[-\hat X]\,s_\lambda[\hat Y]
 =(-1)^n\sum_{\lambda\vdash n}s_{\lambda'}[\hat X]\,s_\lambda[\hat Y],
\end{split}
\end{equation*}
that is, the dual Cauchy identity
\begin{equation}
 \lambda^n(X)=\sum_{\lambda\vdash n}s_{\lambda'}[\hat X]\,s_\lambda[\hat Y].
 \label{eq:dualCauchy}
\end{equation}
The two statistics differ only in the pair of shapes carried by the two factors,
\begin{equation}
 \text{bosons}:(\lambda,\lambda),
 \qquad
 \text{fermions}:(\lambda',\lambda),
 \label{eq:shapetrans}
\end{equation}
and since $\lambda^n[\hat X\hat Y]=\lambda^n[\hat Y\hat X]$ the conjugation may be placed on either factor. This is the parity law of Sec.\ref{sec:establish} in the Schur basis. Conjugating one label transmutes the statistics, conjugating both restores it.

\subsection{Descent sets and the two-alphabet theorem}
\label{ssec:descent}

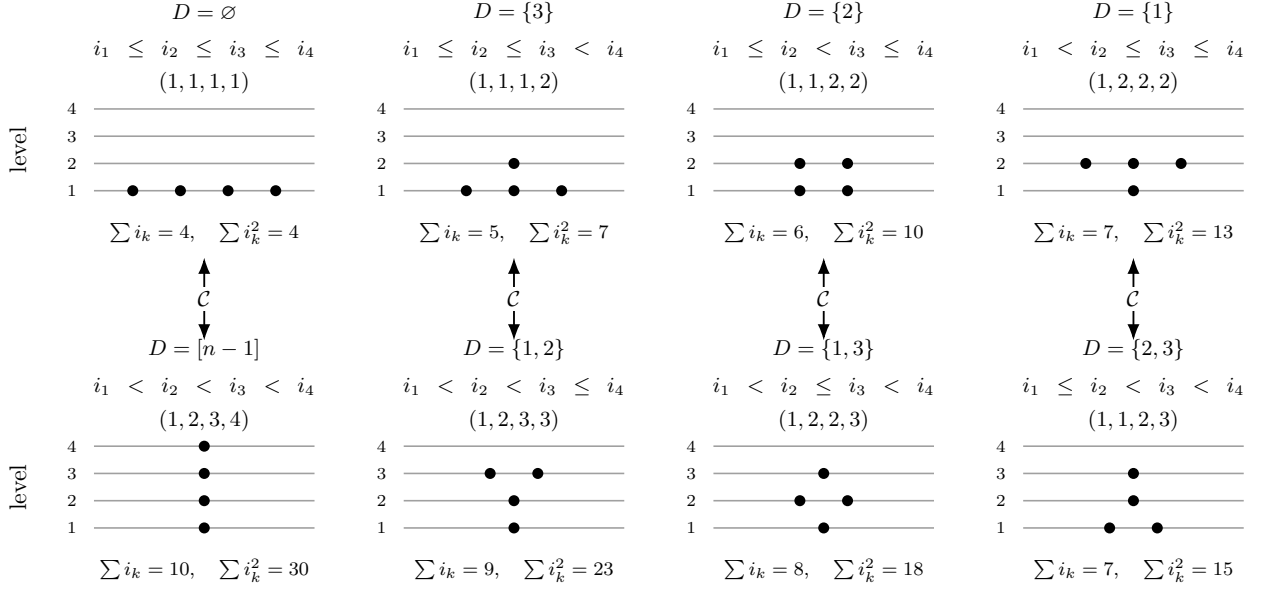
\begin{figure*}
\centering
\begin{tikzpicture}[>=Latex,scale=0.9,every node/.style={transform shape}]
\def\R{-4.95}
\dpan{0cm}{0cm}{$D=\varnothing$}{$\leq$}{$\leq$}{$\leq$}{$(1,1,1,1)$}{4}{4}
\begin{scope}
 \fill (0.45,0) circle (2.3pt);\fill (1.15,0) circle (2.3pt);
 \fill (1.85,0) circle (2.3pt);\fill (2.55,0) circle (2.3pt);\end{scope}
\dpan{4.55cm}{0cm}{$D=\{3\}$}{$\leq$}{$\leq$}{$<$}{$(1,1,1,2)$}{5}{7}
\begin{scope}[xshift=4.55cm]
 \fill (0.8,0) circle (2.3pt);\fill (1.5,0) circle (2.3pt);
 \fill (2.2,0) circle (2.3pt);\fill (1.5,0.4) circle (2.3pt);\end{scope}
\dpan{9.1cm}{0cm}{$D=\{2\}$}{$\leq$}{$<$}{$\leq$}{$(1,1,2,2)$}{6}{10}
\begin{scope}[xshift=9.1cm]
 \fill (1.15,0) circle (2.3pt);\fill (1.85,0) circle (2.3pt);
 \fill (1.15,0.4) circle (2.3pt);\fill (1.85,0.4) circle (2.3pt);\end{scope}
\dpan{13.65cm}{0cm}{$D=\{1\}$}{$<$}{$\leq$}{$\leq$}{$(1,2,2,2)$}{7}{13}
\begin{scope}[xshift=13.65cm]
 \fill (1.5,0) circle (2.3pt);\fill (0.8,0.4) circle (2.3pt);
 \fill (1.5,0.4) circle (2.3pt);\fill (2.2,0.4) circle (2.3pt);\end{scope}
\dpan{0cm}{\R cm}{$D=[n-1]$}{$<$}{$<$}{$<$}{$(1,2,3,4)$}{10}{30}
\begin{scope}[yshift=\R cm]
 \fill (1.5,0) circle (2.3pt);\fill (1.5,0.4) circle (2.3pt);
 \fill (1.5,0.8) circle (2.3pt);\fill (1.5,1.2) circle (2.3pt);\end{scope}
\dpan{4.55cm}{\R cm}{$D=\{1,2\}$}{$<$}{$<$}{$\leq$}{$(1,2,3,3)$}{9}{23}
\begin{scope}[xshift=4.55cm,yshift=\R cm]
 \fill (1.5,0) circle (2.3pt);\fill (1.5,0.4) circle (2.3pt);
 \fill (1.15,0.8) circle (2.3pt);\fill (1.85,0.8) circle (2.3pt);\end{scope}
\dpan{9.1cm}{\R cm}{$D=\{1,3\}$}{$<$}{$\leq$}{$<$}{$(1,2,2,3)$}{8}{18}
\begin{scope}[xshift=9.1cm,yshift=\R cm]
 \fill (1.5,0) circle (2.3pt);\fill (1.15,0.4) circle (2.3pt);
 \fill (1.85,0.4) circle (2.3pt);\fill (1.5,0.8) circle (2.3pt);\end{scope}
\dpan{13.65cm}{\R cm}{$D=\{2,3\}$}{$\leq$}{$<$}{$<$}{$(1,1,2,3)$}{7}{15}
\begin{scope}[xshift=13.65cm,yshift=\R cm]
 \fill (1.15,0) circle (2.3pt);\fill (1.85,0) circle (2.3pt);
 \fill (1.5,0.4) circle (2.3pt);\fill (1.5,0.8) circle (2.3pt);\end{scope}
\node[font=\normalsize,rotate=90,anchor=south] at (-1.00,0.6) {level};
\node[font=\normalsize,rotate=90,anchor=south] at (-1.00,{0.6+\R}) {level};
\foreach \x in {1.5,6.05,10.6,15.15}{
  \draw[<->,line width=0.7pt] (\x,-0.98) -- (\x,-2.20);
  \node[fill=white,inner sep=1.8pt,font=\small] at (\x,-1.59) {$\mathcal{C}$};}
\end{tikzpicture}
\caption{The eight descent sets for $n=4$, arranged in complementary pairs. Each panel shows the lowest-energy configuration compatible with its constraint set in \eqref{eq:fundamental}, together with the total energies for a linear spectrum, $\sum_k i_k$, and a quadratic spectrum, $\sum_k i_k^2$. Complementation $\mathcal{C}$ exchanges the two rows and reflects the level-index sum, so for a linear spectrum, every column gives $E(D)+E(D^{c})=2n+\binom{n}{2}=14$, independently of $D$. This constancy is what makes the reflection of Appendix~\ref{sec:highharmonic} work, since complementing all $d$ descent sets of a product then sends $\sum_r E(D_r)$ to $d\,[2n+\binom{n}{2}]-\sum_r E(D_r)$, a constant minus the original and hence an overall power of $b$. The $2dn$ here comes from setting the ground level energy to 1 rather than 0. The column sums for $\sum_k i_k^2$ are $34,30,28,28$, with no constant to extract, and therefore no such reflection exists once the levels are not linear in the index, such as for the rectangular well of Example~\ref{ex:well}.}
\label{fig:descent}
\end{figure*}

Equation \eqref{eq:shapetrans} distinguishes the two statistics by a partition label, which is a global object. In the basis of fundamental quasisymmetric functions $F_{n,D}$, the same distinction is encoded positionwise by a descent set. For $[n-1] =\{1,2,\ldots,n-1\}$ and $D\subseteq[n-1]$, one has \cite{StanleyEC2}
\begin{equation}
 F_{n,D}[\hat X]=
 \sum_{\substack{1\le i_1\le\cdots\le i_n\\ r\in D\,\Rightarrow\,i_r<i_{r+1}}}
 \hat x_{i_1}\cdots\hat x_{i_n}.
 \label{eq:fundamental}
\end{equation}
The summation runs over weakly increasing sequences of $n$ one-particle state indices, $i_1\le i_2\le\cdots\le i_n$. Such a sequence records an $n$-particle occupation configuration in the factor $\hat X$. Repeated indices represent several particles occupying the same state, while a strict increase marks a transition to a different state with equal or higher energy. If some $\hat x_a$ has degeneracy $g_a$, we account for it by setting $\hat x_{a},\ldots,\hat x_{a+g_a-1}$ to the same Boltzmann weight as $\hat x_a$, so that the ordered sequence of indices still distinguishes the $g_a$ degenerate one-particle states. The subset $D$ specifies the positions at which equality is forbidden. Thus, if $r\in D$, the condition $i_r<i_{r+1}$ prevents the $r$-th and $(r+1)$-th entries of the ordered sequence from referring to the same one-particle state. In this sense, $D$ is a local occupation constraint imposed along the ordered list of particle labels. Fig.\ref{fig:descent} shows the eight constraint sets for $n=4$ and their lowest-energy configurations.

The two extreme choices recover the familiar statistics in a single factor. For $D=\varnothing$, no strict inequalities are imposed, so arbitrary repetitions are allowed, as for bosonic occupation. For $D=[n-1]$, every adjacent pair must be distinct, giving $i_1<i_2<\cdots<i_n$, and hence at most one particle may occupy each one-particle state, as required by Pauli exclusion. Intermediate subsets $D$ impose exclusion only at selected adjacent positions and therefore provide a descent-resolved interpolation between unrestricted and fully exclusive occupation.

Schur functions expand in this basis over standard tableaux \cite[Thm.~7.19.7]{StanleyEC2},
\begin{equation}
 s_\lambda[\hat X]=\sum_{T\in\SYT(\lambda)}F_{n,\Des(T)}[\hat X],
 \label{eq:stdschur}
\end{equation}
where $\Des(T)$ is the set of $r$ such that $r+1$ lies in a lower row of $T$ than $r$, and $L_{\mathrm{co}(T)}=F_{n,\Des(T)}$ in Stanley's notation. Inserting \eqref{eq:stdschur} into both factors of the Cauchy identity \eqref{eq:Cauchy}, and into the dual Cauchy identity \eqref{eq:dualCauchy}, leaves in each case a sum over ordered pairs of standard tableaux of a common shape. The RSK correspondence is a bijection between such pairs and permutations, $\pi\leftrightarrow(P(\pi),Q(\pi))$, under which $\Des(P(\pi))=\Des(\pi^{-1})$ and $\Des(Q(\pi))=\Des(\pi)$ \cite{StanleyEC2}. It therefore replaces the sum over shapes and tableaux by a single sum over $S_n$ and eliminates the shape label altogether, giving \cite[Thm.7.23.2]{StanleyEC2}
\begin{equation}
\begin{split}
 \sum_{\lambda\vdash n}s_\lambda[\hat X]s_\lambda[\hat Y]
 &=\sum_{\pi\in S_n}F_{n,\Des(\pi^{-1})}[\hat X]\,F_{n,\Des(\pi)}[\hat Y],
 \\
 \sum_{\lambda\vdash n}s_{\lambda'}[\hat X]s_{\lambda}[\hat Y]
 &=\sum_{\pi\in S_n}F_{n,\Des(\pi^{-1})^c}[\hat X]\,F_{n,\Des(\pi)}[\hat Y].
 \label{eq:stanley}
\end{split}
\end{equation}
The shape label is thereby replaced by the pair of permutation statistics
\begin{equation}
\begin{split}
 &\Des(\pi)=\{r\in[n-1]:\pi(r)>\pi(r+1)\},
 \\
 &\Des(\pi)^c=[n-1]\setminus\Des(\pi),
 \label{eq:despi}
\end{split}
\end{equation}
and the complement in the second line of \eqref{eq:stanley} is the image of the conjugated shape. Each $\pi$ contributes a pair of constraint sets, one to each factor, and the two statistics are distinguished by whether that pair is $(\Des(\pi^{-1}),\Des(\pi))$ or $(\Des(\pi^{-1}),\Des(\pi)^c)$.

\begin{theorem}[Two-Alphabet descent transmutation]
\label{thm:descent}
For arbitrary alphabets $\hat X,\hat Y$ and every $n\ge0$,
\begin{equation}
\begin{split}
 \sigma^n(\hat X\hat Y)&=\sum_{\pi\in S_n}
 F_{n,\Des(\pi^{-1})}[\hat X]\;F_{n,\Des(\pi)}[\hat Y],\\
 \lambda^n(\hat X\hat Y)&=\sum_{\pi\in S_n}
 F_{n,\Des(\pi^{-1})}[\hat X]\;F_{n,\Des(\pi)^c}[\hat Y]
 \\ &=\sum_{\pi\in S_n}
 F_{n,\Des(\pi^{-1})^c}[\hat X]\;F_{n,\Des(\pi)}[\hat Y].
\end{split}
 \label{eq:twofact-boseferm-descent}
\end{equation}
Thus, the Bose-Fermi transmutation is a complementation of the descent set of exactly one alphabet.
\end{theorem}
The Bose-Fermi switch is therefore represented by a $\ZZ_2$-involution acting on either factor (not both). Writing $\mathcal C$ for the involution $F_{n,D}\mapsto F_{n,D^c}$, whose restriction to symmetric functions sends $s_\lambda\mapsto s_{\lambda'}$, and $\mathcal C_{\hat X},\mathcal C_{\hat Y}$ for its action on the corresponding factor,
\begin{equation}
\begin{split}
 &\mathcal C_{\hat X}\,\sigma^n(\hat X\hat Y)
 =\mathcal C_{\hat Y}\,\sigma^n(\hat X\hat Y)
 =\lambda^n(\hat X\hat Y),
 \\
 &\mathcal C_{\hat X}\mathcal C_{\hat Y}\,\sigma^n(\hat X\hat Y)
 =\sigma^n(\hat X\hat Y),
 \label{eq:Z2descent}
\end{split}
\end{equation}
which is the parity law of Sec.\ref{sec:establish} resolved to the level of individual terms. The transmutation is the exchange
\begin{equation}
 \bigl[r\in D\Rightarrow i_r<i_{r+1}\bigr]
 \;\longleftrightarrow\;
 \bigl[r\notin D\Rightarrow i_r<i_{r+1}\bigr]
 \label{eq:strictness}
\end{equation}
in one factor. Thus partition conjugation gives the global, shape-theoretic description of the transmutation, while descent-set complementation gives its occupation-wise realization by toggling, at each adjacent pair, whether strict inequality is required.

\begin{example}[Two-dimensional harmonic trap]
The one-dimensional oscillator has nondegenerate levels $\epsilon_k=k+\tfrac12$, so $z_1^{(1)}=b^{1/2}/(1-b)$ with $b=e^{-\beta}$ in units $\hbar\omega=1$, and removing the zero point leaves the geometric alphabet $\hat X=(1-b)^{-1}$. In two dimensions the energies add across the Cartesian directions, so $X=\hat X\hat Y=(1-b)^{-2}$ with both factors equal. Neither factor is degenerate and the shell degeneracies of the trap come entirely from the collection \eqref{eq:productdegeneracy}. Within any sum over $\operatorname{SYT}(\lambda)$, principal specialization sends each quasisymmetric function to \cite[Lem.~7.19.10 - 7.19.11]{StanleyEC2},
\begin{equation}
\begin{split}
 &F_{n,D}[\hat X]=\frac{b^{\maj(D)}}{(b)_n},
 \\
 \maj(D)=&\sum_{r\in D}r,
 \qquad
 (b)_n=\prod_{k=1}^{n}(1-b^k).
 \label{eq:principalspec}
\end{split}
\end{equation}
The descent set is compressed to its major index and the complement acts by reflection,
\begin{equation}
 \maj(D^c)=\binom n2-\maj(D).
 \label{eq:majcomp}
\end{equation}
Writing $\maj(\pi)=\maj(\Des(\pi))$ and inserting \eqref{eq:principalspec} and \eqref{eq:majcomp} into Theorem~\ref{thm:descent} gives
\begin{equation}
\begin{split}
 \sigma^n(X)&=\frac{1}{(b)_n^{2}}
 \sum_{\pi\in S_n}b^{\,\maj(\pi^{-1})+\maj(\pi)},\\
 \lambda^n(X)&=\frac{b^{\binom n2}}{(b)_n^{2}}
 \sum_{\pi\in S_n}b^{\,\maj(\pi^{-1})-\maj(\pi)}.
\end{split}
 \label{eq:harmonicmaj}
\end{equation}
The single complement has turned the sum of the two major indices into their difference. A second complement would reflect the other exponent as well and return \eqref{eq:harmonicmaj} to its first line. By Foata -Sch\"utzenberger\cite{FoataSchutzenberger1978} the pair $(\maj(\pi^{-1}),\maj(\pi))$ is jointly equidistributed with $(\inv(\pi),\maj(\pi))$, and restoring the zero-point factor $b^{nd/2}=b^n$, 
\begin{equation}
\begin{split}
 Z^2_{n,B}=\frac{b^n}{(b)_n^{2}}
 \sum_{\pi\in S_n}b^{\,\maj(\pi)+\inv(\pi)},
 \\
 Z^2_{n,F}=\frac{b^{\binom{n+1}2}}{(b)_n^{2}}
 \sum_{\pi\in S_n}b^{\,\maj(\pi)-\inv(\pi)}.
 \label{eq:harmonicclosed}
\end{split}
\end{equation}
The second being (10) of Ref.\cite{ChaudharyValenzuelaChin2026}. The two statistics differ only in the sign of $\inv$, and that sign is the reflection \eqref{eq:majcomp} applied to one factor. Both sums are subtraction-free and can be evaluated in $O(n^2)$ time by the Baxter-Zeilberger recursion\cite{BaxterZeilberger2011}. The compression \eqref{eq:principalspec} is available only because the geometric factors provide the principal specialization.
\end{example}

\subsection{Descent Set as Occupation matrices}

The same two identities have an even more direct reading in terms of occupation numbers, obtained by running RSK in the other direction, making the complement operation more concrete. Let $M=(M_{ij})$ be a finite-support matrix with $\sum_{i,j}M_{ij} = n$, its rows and columns indexed by the letters of $\hat X$ and $\hat Y$ counted with degeneracy, so that $M_{ij}$ is the occupation of the product state indexed by $(i,j)$, whose energy is $\epsilon_i+\eta_j$. With row and column sums $\mu_i=\sum_jM_{ij}$ and $\nu_j=\sum_iM_{ij}$ and weight $\mathrm{wt}(M)=\prod_{i,j}(\hat x_i\hat y_j)^{M_{ij}}=\prod_i\hat x_i^{\mu_i}\prod_j\hat y_j^{\nu_j}$, RSK is a bijection between such matrices and pairs of semistandard tableaux of a common shape \cite{Knuth1970, StanleyEC2},
\begin{equation}
 M\longleftrightarrow
 \bigsqcup_{\lambda\vdash n}\SSYT(\lambda,\mu)\times\SSYT(\lambda,\nu),
 \label{eq:RSKM}
\end{equation}
while dual RSK restricts the entries to a binary matrix $B$, i.e. $B_{ij}\in\{0,1\}$, and conjugates one shape \cite{Knuth1970, StanleyEC2},
\begin{equation}
 B\longleftrightarrow
 \bigsqcup_{\lambda\vdash n}\SSYT(\lambda',\mu)\times\SSYT(\lambda,\nu).
 \label{eq:RSKB}
\end{equation}
Expanding each Schur function over semistandard tableaux
\cite[Def. 7.10.1]{StanleyEC2},
\begin{equation}
 s_\lambda[\hat X]=\sum_{T\in\SSYT(\lambda)}\hat x^{T},
 \label{eq:kostka}
\end{equation}
the Cauchy identity \eqref{eq:Cauchy} becomes a sum over pairs of semistandard tableaux of a common shape, weighted by $\hat x^{P}\hat y^{Q}$. RSK \eqref{eq:RSKM} sends such a pair to a matrix $M$ with $\mathrm{wt}(M)=\hat x^{P}\hat y^{Q}$, changing the sum term by term into a sum over matrices. Applying the same expansion to \eqref{eq:dualCauchy}, where dual RSK \eqref{eq:RSKB} restricts the image to binary matrices,
\begin{equation}
\begin{split}
 &\sigma^n(\hat X\hat Y)=\sum_{\substack{M=(M_{ij})_{(i,j)\in I_{\hat X}\times I_{\hat Y}},\\ M_{ij} \in \NN_0,\, \sum_{i,j} M_{ij} = n}}\mathrm{wt}(M),
 \\
 &\lambda^n(\hat X\hat Y)=\sum_{\substack{B=(B_{ij})_{(i,j)\in I_{\hat X}\times I_{\hat Y}},\\ B_{ij} \in \{0,1\},\, \sum_{i,j} B_{ij} = n}}\mathrm{wt}(B).
 \label{eq:Moccupation}
\end{split}
\end{equation}
Here $I_{\hat X}$ and $I_{\hat Y}$ denote the index sets of the letters of $\hat X$ and $\hat Y$, so that $I_{\hat X}\times I_{\hat Y}$ specifies the allowed product states. The sets $\NN_0=\{0,1,2,\ldots\}$ and $\{0,1\}$ specify their possible bosonic and fermionic occupation numbers, respectively. The binary restriction is Pauli exclusion written in the product-state basis. Under \eqref{eq:RSKB} it becomes the conjugation of one shape in \eqref{eq:shapetrans}, and after standardization the complement of one descent
set.

\begin{example}[Rectangular infinite well]
\label{ex:well}
For a one-dimensional well the levels are $\epsilon_k=\Delta k^2$, $k\ge1$, all nondegenerate, so
\begin{equation}
 \hat X(b_x)=\{b_x^{1^2},b_x^{2^2},b_x^{3^2},\ldots\},
 \qquad
 b_x=e^{-\beta\Delta_x},
\end{equation}
and likewise $\hat Y(b_y)$. In two dimensions the energies $\Delta_xk^2+\Delta_yl^2$ separate, so $X=\hat X(b_x)\hat Y(b_y)$. The shell degeneracies are now the number of representations of an integer as a sum of two squares.

Because the weight of an occupation matrix depends on it only through the row and
column sums $\mu_k=\sum_lM_{kl}$ and $\nu_l=\sum_kM_{kl}$,
\begin{equation*}
\begin{split}
 \mathrm{wt}(M)&=\prod_k\bigl(b_x^{k^2}\bigr)^{\mu_k}\prod_l\bigl(b_y^{l^2}\bigr)^{\nu_l}\\
 &=b_x^{\sum_kk^2\mu_k}\,b_y^{\sum_ll^2\nu_l}
 \;\xrightarrow{\;b_x=b_y=b\;}\;
 b^{\sum_ii^2(\mu_i+\nu_i)},
 \label{eq:wellweight}
\end{split}
\end{equation*}
and \eqref{eq:Moccupation} becomes
\begin{equation}
\begin{split}
 \sigma^n(X)=\sum_{\substack{M=(M_{ij})_{(i,j)\in \NN^2},\\ M_{ij} \in \NN_0,\, \sum_{i,j} M_{ij} = n}}b^{\sum_kk^2(\mu_k+\nu_k)},
 \\
 \lambda^n(X)=\sum_{\substack{B=(B_{ij})_{(i,j)\in \NN^2},\\ B_{ij} \in \{0,1\},\, \sum_{i,j} B_{ij} = n}}b^{\sum_kk^2(\mu_k+\nu_k)},
 \label{eq:wellocc}
\end{split}
\end{equation}
the two differing only in the range of the entries. The principal specialization \eqref{eq:principalspec} requires the exponents to be linear in the level index.
\end{example}

\section{Physical realizations of two-alphabet spectra}
\label{sec:realizations}

So far, the examples we have shown for two-alphabet spectra consist only of two-dimensional systems that are separable in those two dimensions. But the product $X=\hat X\hat Y$ was introduced by separating a spectrum into two additive sectors, and nothing in \eqref{eq:twoalph} refers to what those sectors are. Whenever the one-particle energies satisfy
\begin{equation}
 E_{ij}=\epsilon_i+\eta_j+E_0,
 \label{eq:additive}
\end{equation}
the alphabet is $e^{-\beta E_0}\hat X\hat Y$, the constant contributing an overall factor of $e^{-n\beta E_0}$ to every $n$-particle partition function and dropping out of everything below. Now we illustrate cases where this two-alphabet spectra originates not from the dimension separability but other factorizations.

\subsection{Orbital and internal degrees of freedom}

Let $\hat X$ collect the orbital weights $e^{-\beta\epsilon_i}$ and $\hat Y$ the internal weights $e^{-\beta\Delta_j}$ of a spin or other internal multiplet, the internal shift being independent of the orbital label. An unsplit multiplet of $g$ internal states is the alphabet with a single letter of degeneracy $g$ at zero energy,
\begin{equation}
 \hat Y=\underbrace{1+\cdots+1}_{g}\;\equiv\;1_g,
\end{equation}
which is repetition of letters in the sense of \eqref{eq:generalalphabet}. The binary constraint \eqref{eq:Moccupation} applies to the complete product state $(i,j)$, so an orbital level $i$ admits up to $g$ fermions, one per internal state. The same statement appears in the Schur expansion through the hook-content formula
\cite[Cor.~7.21.4]{StanleyEC2},
\begin{equation}
s_\lambda[1_g]=\prod_{(r,c)\in\lambda}\frac{g+c-r}{h(r,c)},
 \label{eq:hookcontent}
\end{equation}
which vanishes when $\lambda$ has more than $g$ rows. In \eqref{eq:dualCauchy}
the internal factor carries $\lambda$ and the orbital factor carries $\lambda'$, so $s_\lambda[1_g]=0$ whenever the orbital shape has width $\lambda'_1>g$. The bound on the orbital Young diagram is therefore the representation-theoretic form of the capacity bound.

For $n=2$ and $g=2$, with $s_{(2)}[1_2]=3$ and $s_{(1,1)}[1_2]=1$,
\begin{equation}
\begin{split}
 \sigma^2(\hat X1_2)=3\,s_{(2)}[\hat X]+s_{(1,1)}[\hat X],
 \\
 \lambda^2(\hat X1_2)=s_{(2)}[\hat X]+3\,s_{(1,1)}[\hat X],
 \label{eq:twospin}
\end{split}
\end{equation}
the triplet pairing with the symmetric orbital shape in the first case and with the antisymmetric one in the second. Passing between the two lines is $\mathcal C_{\hat Y}$ applied before specializing $\hat Y=1_2$. For $n=3$, $s_{(1,1,1)}[1_2]=0$, so
\begin{equation}
 \lambda^3(\hat X1_2)=4\,s_{(1,1,1)}[\hat X]+2\,s_{(2,1)}[\hat X],
\end{equation}
and the fully symmetric orbital shape $(3)$ cannot occur for three spin-$\tfrac12$ fermions. A Zeeman or hyperfine splitting replaces $1_g$ by $\hat Y=\sum_je^{-\beta\Delta_j}$ and changes nothing structural, provided the splitting remains independent of the orbital label.

The unsplit $g$-fold degree of freedom can also be treated as direct sum rather than tensor product by,
\begin{equation}
\mathcal H_{\hat X}\otimes\mathbb C^g
\simeq
\bigoplus_{i=1}^g \mathcal H_{\hat X},
\end{equation}
giving the alphabet separation as a sum instead, $X=\sum_{i=1}^g \hat X = 1_g \hat X$.

\subsection{Circular modes of the Fock-Darwin oscillator}
A charged two-dimensional oscillator in a perpendicular field\cite{Fock1928,Darwin1931} has the spectrum\cite{DRIGHOFILHO2017101}
\begin{equation}
 E_{n_+n_-}=\omega_+\Bigl(n_++\tfrac12\Bigr)+\omega_-\Bigl(n_-+\tfrac12\Bigr),
 \qquad n_\pm\in\NN,
\end{equation}
where $\Omega=\sqrt{\omega_0^2+\omega_c^2/4}$ and $\omega_\pm=\Omega\pm\omega_c/2$. Removing the zero point
\begin{equation}
 \hat X=(1-b_+)^{-1},
 \quad
 \hat Y=(1-b_-)^{-1},
 \quad
 b_\pm=e^{-\beta\omega_\pm}.
 \label{eq:fockdarwin}
\end{equation}
The factorization is with respect to the circular normal modes, not the Cartesian axes in which the Hamiltonian is naturally written, and the $\ZZ_2$ acts on either mode. Both factors are geometric, so the principal specialization of Theorem~\ref{thm:descent} applies with two independent parameters and produces a bivariate refinement of the Mahonian form \eqref{eq:harmonicclosed}, with the isotropic case $b_+=b_-$ ($\omega_c = 0$) recovering Ref.\cite{ChaudharyValenzuelaChin2026}.

\section{Arbitrary Product Spectra}
\label{sec:kronecker}

Let $X=A_1A_2\cdots A_d$. We first restate two standard facts. First, Adams operations are multiplicative on products $p_\rho[X]=\prod_r p_\rho[A_r]$, and second, the power sums expand in the Schur basis through the characters of $S_n$ \cite[I.(7.8)]{Macdonald1995},
\begin{equation}
 p_\rho[A]=\sum_{\lambda\vdash n}\chi^\lambda_\rho\,s_\lambda[A].
 \label{eq:frobenius}
\end{equation}

In the power-sum basis \cite[I.(2.14$'$)]{Macdonald1995}
\begin{equation}
 \sigma^n(X)=\sum_{\rho\vdash n}\frac{p_\rho[X]}{z_\rho},
 \qquad
 \lambda^n(X)=\sum_{\rho\vdash n}\varepsilon_\rho\frac{p_\rho[X]}{z_\rho},
 \label{eq:cycleexp}
\end{equation}
where $\varepsilon_\rho=(-1)^{n-l(\rho)}$, $z_\rho=\prod_ii^{m_i}m_i!$ for $\rho=(1^{m_1}2^{m_2}\cdots)$ and $l(\rho)=\sum_im_i$. Both sums are over cycle types, and $n!/z_\rho$ is the number of permutations of type $\rho$, so they are averages over $S_n$, giving
\begin{equation}
\begin{split}
 \sigma^n(X)&=\frac1{n!}\sum_{\pi\in S_n}p_{\rho(\pi)}[X],
 \\
 \lambda^n(X)&=\frac1{n!}\sum_{\pi\in S_n}\sgn(\pi)\,p_{\rho(\pi)}[X],
\end{split}
\end{equation}
where $\rho(\pi)$ is the cycle type of $\pi$. The $\sgn(\pi)$ comes from the fact that the permutation of type $\rho$ is a product of $n-l(\rho)$ transpositions, so $\varepsilon_\rho=\sgn(\pi)$. Expanding this using \eqref{eq:frobenius} on the multiplicative factors, we have
\begin{equation}
\begin{split}
\sigma^n(X)=\sum_{\boldsymbol{\lambda}}g^{+}_d(\boldsymbol{\lambda})\prod_{r=1}^ds_{\lambda_r}[A_r],
 \\
 \lambda^n(X)=\sum_{\boldsymbol{\lambda}}g^{-}_d(\boldsymbol{\lambda})\prod_{r=1}^ds_{\lambda_r}[A_r],
 \label{eq:dCauchy}
\end{split}
\end{equation}
summed over $d$-tuples of partitions of $n$. Here the generalized Kronecker coefficients\cite{highestweightvectorstensors} come from collecting the coefficients of $\prod_rs_{\lambda_r}[A_r]$,
\begin{equation*}
\begin{split}
 g^{+}_d(\boldsymbol{\lambda})
&=\bigl\langle\chi^{\lambda_1}\cdots\chi^{\lambda_d},\,1\bigr\rangle_{S_n} = \frac1{n!}\sum_{\pi\in S_n}\prod_{r=1}^d\chi^{\lambda_r}(\pi),
 \\
 g^{-}_d(\boldsymbol{\lambda})
&=\bigl\langle\chi^{\lambda_1}\cdots\chi^{\lambda_d},\,\sgn\bigr\rangle_{S_n} = \frac1{n!}\sum_{\pi\in S_n}\sgn(\pi)\prod_{r=1}^d\chi^{\lambda_r}(\pi).
\end{split}
\end{equation*}
Since $[\lambda']=[\lambda]\otimes\sgnrep$, 
conjugating any one label exchanges the two multiplicities,
\begin{equation}
 g^{-}_d(\lambda_1,\ldots,\lambda_d)
 =g^{+}_d(\lambda_1,\ldots,\lambda_j',\ldots,\lambda_d),
 \label{eq:gexchange}
\end{equation}
for any $j$, and conjugating two of them restores $g^+_d$. What is special about $d=2$ is that the multiplicities are trivial\cite{highestweightvectorstensors},
\begin{equation}
g^{+}_2(\lambda,\mu)=\delta_{\lambda\mu},\qquad
g^{-}_2(\lambda,\mu)=\delta_{\lambda\mu'},
 \label{eq:g2}
\end{equation}
so \eqref{eq:dCauchy} collapses to the Cauchy and dual Cauchy identities \eqref{eq:Cauchy} and \eqref{eq:dualCauchy}. 

Inserting the standardization expansion \eqref{eq:stdschur} into \eqref{eq:dCauchy} resolves each factor into descent sets,
\begin{equation}
\begin{split}
 \sigma^n(X)&=\sum_{\boldsymbol\lambda}g^{+}_d(\boldsymbol\lambda)
 \,\prod_{r=1}^{d}\quad\sum_{T_r\in\SYT(\lambda_r)}F_{n,\Des(T_r)}[A_r],
 \\
 \lambda^n(X)&=\sum_{\boldsymbol\lambda}g^{-}_d(\boldsymbol\lambda)
 \,\prod_{r=1}^{d}\quad\sum_{T_r\in\SYT(\lambda_r)}F_{n,\Des(T_r)}[A_r],
 \label{eq:dDescent}
\end{split}
\end{equation}

\begin{theorem}[Multi-Alphabet Transmutation]
\label{thm:dparity}
Let $\mathcal C_r$ denote the involution of \eqref{eq:Z2descent} acting on the $r$th alphabet of \eqref{eq:dDescent}, complementing its descent sets $F_{n,D}\mapsto F_{n,D^c}$ and, on schur functions, conjugating its shape $s_{\lambda_r}\mapsto s_{\lambda_r'}$, and let $\mathcal C_S=\prod_{r\in S}\mathcal C_r$ for $S\subseteq[d]$. Then
\begin{equation}
 \mathcal C_S\,\sigma^n(X)=
 \begin{cases}
 \sigma^n(X),&|S|\text{ even},\\
 \lambda^n(X),&|S|\text{ odd},
 \end{cases}
 \label{eq:dparity}
\end{equation}
and likewise with $\sigma^n$ and $\lambda^n$ interchanged. Thus Bose-Fermi transmutation is complementation of any odd number of descent sets.
\end{theorem}

It is easy to see why this is true. Since the sum runs over all $d$-tuples of partitions of $n$ and $\lambda\mapsto\lambda'$ is an involution on them, each label conjugated by $\mathcal C_S$ may be relabelled back, which moves the conjugation onto the multiplicity, $g^+_d(\boldsymbol\lambda)\mapsto g^+_d(\boldsymbol\lambda^S)$ with $\boldsymbol\lambda^S$ the tuple conjugated in $S$. By \eqref{eq:gexchange} each conjugation exchanges $g^+_d$ with $g^-_d$, so only $|S|$ being odd or even matters. Equivalently, the fermionic line of \eqref{eq:dDescent} may be written with the bosonic multiplicity and any odd number of complements,
\begin{equation*}
\begin{split}
 \lambda^n(X)=\sum_{\boldsymbol\lambda}g^{+}_d(\boldsymbol\lambda)
 \prod_{r\in S}\;\sum_{T_r\in\SYT(\lambda_r)}F_{n,\Des(T_r)^c}[A_r] \\
 \times \prod_{r\notin S}\;\sum_{T_r\in\SYT(\lambda_r)}F_{n,\Des(T_r)}[A_r],
 \label{eq:dDescentF}
\end{split}
\end{equation*}
for any $S$ such that $|S|$ is odd.

The occupation representation extends directly to $d$ alphabets. A bosonic configuration is a finite-support $d$-dimensional array $M_{i_1\cdots i_d}\in\mathbb N_0$, while a fermionic configuration is the binary array $B_{i_1\cdots i_d}\in\{0,1\}$, with total occupation $n$. Summing their weights reproduces the generalized-Kronecker expansions above. Thus the matrix picture of Sec.\ref{sec:twofactor} survives in arbitrary dimension at the level of generating functions, even though for $d>2$ there is no general RSK-type bijection resolving each individual array into a unique tuple of tableaux.

For the isotropic harmonic trap, where principal specialization makes complementing all $d$ descent sets equivalent to $b \mapsto b^{-1}$, taking $|S|=d$ recovers the odd-dimensional relation $Z^d_{n,F}(\beta)=(-1)^nZ^d_{n,B}(-\beta)$ of Ref.\cite{SchmidtSchnack1999}, and a symmetric reflection for even dimensions (details in Appendix~\ref{sec:highharmonic}).

\section{Tensor resolutions of a spectral alphabet}
\label{sec:tensor_resolution}

In the previous sections, we assumed some factorization of the Hilbert space with the Hamiltonian acting independently on the factors. In this section, we ask the question in the other direction ``does the spectrum itself determine the factorization?" We will illuminate this using two examples already presented above, the harmonic oscillator and the rectangular infinite well.

We have so far already established that the alphabet $X$ fixes the canonical thermodynamics, i.e. the canonical partition function is invariant under Hilbert space factorization and is insensitive to how that alphabet is realized as a tensor product. A factorization may arise naturally, as in a Hamiltonian separable in spatial dimensions, or more abstractly through positive spectral alphabets that can themselves be associated with independent Hilbert space factors. The question of tensor factorization therefore becomes, at the spectral level, the question of how $X$ can be factored.

After shifting all ground-state energies to zero, let $\Sigma_2(X)$ be the set of two-factor resolutions,
\begin{equation}
\begin{split}
    \Sigma_2(X)= &\{
(A,B):X=AB
\\&A=\Tr_{\mathcal H_{A}}\, e^{-\beta H_{A}},\; B = \Tr_{\mathcal H_{B}}\, e^{-\beta H_{B}}\}\big/\!\sim
\end{split}
\end{equation}

where $(A,B)\sim(B,A)$.  The trivial resolution $(X,1)$ is included. The set $\Sigma_2(X)$ therefore records the positive two-factor tensor resolutions compatible with a fixed thermodynamic alphabet. For our purposes, the set $\Sigma_2(X)$ is sufficient since any three-factor tensor resolution is also a two-factor tensor resolution. Now we determine $\Sigma_2(X)$ for the examples above and show that spectra can differ quite largely in how much tensor structure they permit.

\paragraph{One-dimensional box.}
For a spectral alphabet $X=\sum_j e^{-\beta \epsilon_j}$, let the spectral support be the set of distinct energy levels $\mathcal S(X)=\{\epsilon_j\}$. After shifting the ground state and fixing the natural energy unit, the
one-dimensional box has spectral support
\begin{equation*}
\mathcal S(X)=\{n^2-1:n\ge1\}.
\end{equation*}
Suppose that the spectral alphabet admits a positive factorization $X = X_A X_B$, with the supports $\mathcal S(X_A) = A$ and $\mathcal S(X_B) = B$, then $\mathcal S(X) = A + B =\{a+b:a\in A,\ b\in B\}$. Because $0\in B$, one has $A\subseteq \mathcal S(X)$, and similarly $B\subseteq \mathcal S(X)$.

Assume that both factors are nontrivial and choose $a>0$ in $A$. For every $b\in B$, since $b \in \mathcal S(X)$ and $a+b\in \mathcal S(X)$, there exist integers $n$, $m$ such that $b = n^2-1$ and $a+b=m^2-1$. But since for fixed $a$, there are only finitely many values of $(m,n)$ that satisfy $m^2 - n^2 = a$, $B$ is a finite set. Similarly if $B$ contains some $b > 0$, the same argument makes $A$ finite. But since $\mathcal S(X)$ is infinite, both $A$ and $B$ cannot be non-trivial at the same time without being finite resulting in a contradiction. Therefore either $A$ or $B$ must be trivial, and
\begin{equation*}
    \Sigma_2(X) = \{(X,1)\}
\end{equation*}

\paragraph{Harmonic oscillator.}
For the shifted harmonic oscillator, $X = (1-b)^{-1}$. Using the binary expansion of nonnegative integers,
\begin{equation*}
\frac{1}{1-b}
=
\prod_{j\ge0}\left(1+b^{2^j}\right).
\end{equation*}
Hence, for any subset $S\subseteq\mathbb Z_{\ge0}$,
\begin{equation*}
A_S=\prod_{j\in S}\left(1+b^{2^j}\right),
\qquad
B_S=\prod_{j\notin S}\left(1+b^{2^j}\right),
\end{equation*}
gives a positive factorization $X = A_S B_S$. Since there are infinitely many choices of $S$ up to complementation,
\begin{equation*}
|\Sigma_2(X)|=\infty.
\end{equation*}
This is another consequence of the linear spectrum, alongside the principal specialization used in Sec.\ref{sec:twofactor}. The same equal level spacing that makes each factor geometric also lets the one-particle Hilbert space be split into independent subsystems in infinitely many inequivalent ways, and that freedom is what leaves room to look for a factorization in which the problem becomes tractable. A rigid spectrum such as the box offers no such choice.

\paragraph{$m$-level internal degree of freedom.}
Consider an equally spaced $m$-level spectrum
\[
X_m=1+b+\cdots+b^{m-1},
\qquad
b=e^{-\beta\Delta}.
\]
Whenever $m=kl$, one has the positive factorization
\begin{equation*}
X_m
=
\left(\sum_{r=0}^{k-1}b^r\right)
\left(\sum_{s=0}^{l-1}b^{ks}\right).
\end{equation*}
This corresponds to the Hilbert-space resolution $\mathbb C^m\simeq\mathbb C^k\otimes\mathbb C^l$, with $H = \Delta N_k\otimes I + k\Delta I\otimes N_l$, $N_k=\operatorname{diag}(0,1,\ldots,k-1).$

Since $X_m$ is a finite polynomial, it admits only finitely many factorizations into positive spectral polynomials, and hence $|\Sigma_2(X_m)|<\infty$.

For example, for $m=6$, $X_6(q)=1+q+\cdots+q^5$. Up to interchange of the two factors,
\begin{equation*}
\begin{split}
\Sigma_2(X_6)
=
\{
&(X_6,1),\,
(1+q,\;1+q^2+q^4),\\
&(1+q+q^2,\;1+q^3)
\}.
\end{split}
\end{equation*}
Hence, $|\Sigma_2(X_6)|=3$. The two nontrivial resolutions both correspond to a $2\times3$ factorization of the six-dimensional Hilbert space, but assign different spectra to the two factors. Thus $\Sigma_2(X)$ contains more information than the possible factorizations of $\dim\mathcal H$ alone.
\\

The examples therefore show the three possibilities
\begin{equation}
|\Sigma_2(X)|=1,\quad
1<|\Sigma_2(X)|<\infty,\quad
|\Sigma_2(X)|=\infty,
\end{equation}
realized here by the one-dimensional box, the finite $m$-level spectrum, and the harmonic oscillator, respectively.

We did not find a natural infinite spectrum with a finite but nontrivial resolution set without imposing additional restrictions. An example would be a $d$-dimensional incommensurate rectangular well, i.e. the energy scales in each direction are linearly independent over $\mathbb Q$, with $|\Sigma_2|=2^{d-1}$.

\section{Operator-kernel extension}
\label{sec:operator-extension}

The previous sections were formulated in terms of canonical partition functions, for
which the one-particle object is the imaginary-time kernel $K_H(\beta)=e^{-\beta H_1}$. However, the construction can be generalized to any observable.

Let $O$ be a self-adjoint operator on the one-particle Hilbert space
$\mathcal H_1$ with eigenvalues $o_a$ of multiplicity $g_a$, such that the kernel
\begin{equation}
K_O(\alpha)=e^{-\alpha O}
\label{eq:app-observable-kernel}
\end{equation}
is a trace class for $\alpha \in \mathbb R_{> 0}$ (ensuring convergence), and let
\begin{equation}
X_O(\alpha)=\sum_a g_a\,e^{-\alpha o_a}
\label{eq:app-observable-alphabet}
\end{equation}
be the corresponding spectral alphabet. Furthermore, for $n$ particles, the kernel on $\mathcal H_n=\mathcal H_1^{\otimes n}$ is the tensor power $K_O^{(n)}(\alpha) = K_O(\alpha)^{\otimes n}$, i.e.
\begin{equation}
O^{(n)}=\sum_{j=1}^{n}I^{\otimes(j-1)}\otimes O\otimes I^{\otimes(n-j)}.
\label{eq:app-additive-lift}
\end{equation}
Then its Adams operations are
\begin{equation}
\psi^m\bigl(X_O(\alpha)\bigr)
=\sum_a g_a\,e^{-m\alpha o_a}
=\Tr_{\mathcal H_1}K_O(\alpha)^m.
\label{eq:app-observable-adams}
\end{equation}
Let $U_\pi$ denote the action of $\pi\in S_n$ on $\mathcal H_n$ permuting the
tensor slots, and let the partition $\rho(\pi)$ be its cycle type. Since the same $O$ acts in every slot, $K_O(\alpha)^{\otimes n}$
commutes with every $U_\pi$, we can write the exchange projectors
\begin{equation}
P_B=\frac{1}{n!}\sum_{\pi\in S_n}U_\pi ,
\qquad
P_F=\frac{1}{n!}\sum_{\pi\in S_n}\sgn(\pi)\,U_\pi ,
\label{eq:app-exchange-projectors}
\end{equation}
whose ranges are $\operatorname{Sym}^n\mathcal H_1$ and $\bigwedge^n\mathcal H_1$. Permuting the factors and taking
the trace closes each cycle into a single trace of the corresponding power of
the kernel after relabeling of dummy indices,
\begin{equation*}
\begin{split}
\Tr_{\mathcal H_n}&
\left(U_\pi K_O^{(n)}(\alpha)\right) = \sum_{a_1,\ldots,a_n} \prod_{i=1}^n (K_O(\alpha))_{a_{\pi(i)}a_i} \\&= 
\prod_{\rho_i \in \rho(\pi)} \sum_{a_1,\ldots,a_{\rho_i}}(K_O(\alpha))_{a_1,a_2}\ldots(K_O(\alpha))_{a_{\rho_i}a_1} \\
&=\prod_{\rho_i \in \rho(\pi)} \Tr_{\mathcal H_{1}} K_O(\alpha)^{\rho_i} = \prod_{\rho_i \in \rho(\pi)} \psi^{\rho_i}(X_O(\alpha))
\end{split}
\end{equation*}
Averaging against the symmetric and antisymmetric particle exchanges (bosonic and fermionic) projectors \eqref{eq:app-exchange-projectors} we have
\begin{align}
Q^O_{n,B}(\alpha)&\equiv\Tr\left[P_B\,e^{-\alpha O^{(n)}}\right]
=\sigma^n\bigl(X_O(\alpha)\bigr),
\label{eq:app-bosonic-observable-trace}\\
Q^O_{n,F}(\alpha)&\equiv\Tr\left[P_F\,e^{-\alpha O^{(n)}}\right]
=\lambda^n\bigl(X_O(\alpha)\bigr).
\label{eq:app-fermionic-observable-trace}
\end{align}
For a generic $O$, $Q^O_{n,B/F}(\alpha)$ are spectral generating functions for the operator $O$. The cumulants of $O^{(n)}$ can then be computed by taking log derivatives of
\eqref{eq:app-bosonic-observable-trace} and \eqref{eq:app-fermionic-observable-trace},
\begin{equation}
\kappa_j\bigl(O^{(n)}\bigr)_{B/F}
=(-1)^j\,\partial_\alpha^{\,j}\log Q^O_{n,B/F}(\alpha).
\label{eq:app-cumulants}
\end{equation}
The first two being the mean and the variance,
\begin{equation*}
\begin{split}
-\partial_\alpha\log Q^O_{n,B/F}
&=\bigl\langle O^{(n)}\bigr\rangle_{B/F},
\\
\partial_\alpha^{2}\log Q^O_{n,B/F}
&=\Var_{B/F}\bigl(O^{(n)}\bigr).
\end{split}
\end{equation*}

The factorwise identities extend as well. If the one-particle space decomposes
as $\mathcal H_1=\mathcal H_{\hat X}\otimes\mathcal H_{\hat Y}$ with
\begin{equation}
O=O_{\hat X}\otimes I+I\otimes O_{\hat Y},
\label{eq:app-additive-factor-observable}
\end{equation}
then $e^{-\alpha O}=e^{-\alpha O_{\hat X}}\otimes e^{-\alpha O_{\hat Y}}$, so
the spectral alphabet factors as in \eqref{eq:twofactoralph},
\begin{equation}
X_O(\alpha)=\hat X_O(\alpha)\,\hat Y_O(\alpha).
\label{eq:app-factor-observable-alphabet}
\end{equation}
All Cauchy, dual-Cauchy, descent-set, and transmutation identities of the previous sections consequently hold verbatim for $X_O(\alpha)$, and
likewise for a $d$-fold decomposition and Sec.\ref{sec:kronecker}.

More generally, the exponential form is not required by the symmetric-function identities themselves. For any trace-class one-particle operator $K$, one may define a formal alphabet $X_K$ through
\begin{equation*}
\psi^m(X_K)
:=
\Tr (K^m).
\label{eq:app-general-kernel-alphabet}
\end{equation*}

This extends the calculus to more interesting systems. 
\begin{itemize}
 
\item \emph{Non-commuting sources.} For a one-body observable $A$ with $[A,H_1]\neq0$, the kernel $K(\alpha)=e^{-\beta H_1/2}e^{-\alpha A}e^{-\beta H_1/2}$ is positive and trace class but is not the exponential of any operator whose spectrum is known, so $X_O$ does not exist while $X_K$ does. Since $K(\alpha)$ is still a differentiable family, derivatives with respect to $\alpha$ generate $\langle A^{(n)}\rangle_{B/F}$ and its associated fluctuations.
 
\item \emph{Non-self-adjoint generators.} For open or $\mathcal{PT}$-symmetric one-particle systems, the eigenvalues of $K=e^{-\beta H_1}$ can be complex and $X_K$ can be a formal alphabet with complex letters. All Cauchy, dual-Cauchy, descent-set, and transmutation identities continue to hold, since none of them used positivity of the letters. As a simple example, for a separable harmonic oscillator with $\omega_y=\omega_0-i\gamma/2$, the imaginary part of the many-body resonance energy is proportional to the occupation of the lossy factor. The corresponding decay width therefore reads the same Mahonian statistics that appears in the Hermitian transmutation formulas.
\end{itemize}

\section{Conclusion}

In this work, the single-particle spectrum was treated as an object of interest with its own arithmetic. Encoded as an alphabet $X$ whose Adams operations are the power traces, a tensor factorization of the one-particle Hilbert space is a factorization of $X$, and the bosonic and fermionic statistics differ only in how one factor is read. That difference is a conjugated symmetry label in the Schur basis, a complemented descent set in the quasisymmetric one, and a restricted occupation number in the product-state basis, with an odd number of factors complemented when there are more than two. Moreover, these factors need not be spatial directions, nor is the alphabet restricted to a trace of $\exp(-\beta H_1)$.

Since canonical thermodynamics sees only $X$, the same partition functions may be reached through inequivalent factorizations, and the number of factorizations a spectrum admits varies from one to infinitely many. A rigid spectrum such as the box offers no nontrivial resolution, while the oscillator can be split into independent subsystems in infinitely many ways. This freedom provides room to search for a factorization in which the many-particle problem becomes tractable. In the harmonic case the same equal level spacing that supplies it also makes the factors geometric, admitting the principal specialization that compresses the
general expansions into permutation statistics. The freedom itself is not automatic, and in Sec.\ref{sec:tensor_resolution} we found it only for the oscillator or under additional conditions imposed by hand. Whether other natural spectra share it, and whether they are tractable when they do, would settle how much of the harmonic case's solvability the resolution set accounts for.

The natural extension of this work is to remove the restriction to one-dimensional irreducible representations of $S_n$, the trivial and sign representations, which give $\sigma^n$ and $\lambda^n$ respectively. Higher-dimensional irreducible representations give parastatistics, which have recently drawn renewed interest due to their application in quantum computation and realization of genuine paraparticles not reducible to just bosons or fermions \cite{Wang2025,Wang2026}. An instance of this extension has been established in Ref.\cite{ChinChadPara}, where the $\xi$-interpolated recursion at $\xi=-1/m$ yields the partition function of genuine paraparticles. Its grand partition function is exactly $\lambda_{t/m}[X\,1_m]$, so up to normalization it is the internal multiplet alphabet of Sec.\ref{sec:realizations} at $g=m$. Since the alphabet fixes the thermodynamics but not the statistics that produced it, what remains is to relate alphabets to the underlying $R$-matrix exchange.

\appendix
\section{Harmonic Exchange Polynomials and the odd-dimensional reflection}
\label{sec:highharmonic}

Take all $d$ alphabets geometric and equal, $A_r=(1-b)^{-1}$, so that $X=(1-b)^{-d}$ is the alphabet of the isotropic $d$-dimensional harmonic trap without the zero-point energy. Principal specialization \eqref{eq:principalspec} sends each quasisymmetric function to a single power of $b$, so \eqref{eq:dDescent} becomes
\begin{equation}
 \sigma^n(X)=\frac{\mathcal P^{+}_{n,d}(b)}{(b)_n^{\,d}},
 \qquad
 \lambda^n(X)=\frac{\mathcal P^{-}_{n,d}(b)}{(b)_n^{\,d}},
 \label{eq:coreZ}
\end{equation}
with the harmonic exchange polynomials
\begin{equation}
 \mathcal P^{\pm}_{n,d}(b)=\sum_{\boldsymbol\lambda}g^{\pm}_d(\boldsymbol\lambda)
 \prod_{r=1}^{d}\;\sum_{T_r\in\SYT(\lambda_r)}b^{\,\maj(T_r)}.
 \label{eq:core}
\end{equation}

Both are polynomials with nonnegative integer coefficients, since $g^{\pm}_d(\boldsymbol\lambda)$ is the multiplicity of an irreducible representation in $[\lambda_1]\otimes\cdots\otimes[\lambda_d]$ \cite{highestweightvectorstensors}, with degree at most $dM$ and $M=\binom n2$ being the maximal value of $\maj$. For $d=1$, $\mathcal P^{+}_{n,1}=1$ and $\mathcal P^{-}_{n,1}=b^{M}$. For $d=2$, $\mathcal P^{\pm}_{n,2}$ is $\sum_\pi b^{\maj(\pi)\pm\inv(\pi)}$ of \eqref{eq:harmonicclosed} up to the overall power of $b$. Anisotropic traps replace $b$ by one variable per factor throughout, with no change to what follows.

\subsection{Reflection at odd vs even $d$}

The universal statement $\lambda^n(X)=(-1)^n\sigma^n(-X)$ of Sec.\ref{sec:establish} is about negating the alphabet and holds for every $d$. It becomes a statement about the temperature only when $b\mapsto b^{-1}$ realizes that negation, and whether it does is a question of parity. Since $A(b^{-1})=(1-b^{-1})^{-1}=-b\,A(b)$, inverting $b$ negates each factor, so
\begin{equation}
 X(b^{-1})=(-1)^{d}b^{d}\,X(b),
 \label{eq:Xreflect}
\end{equation}
which is $-X$ up to a line element precisely when $d$ is odd. For even $d$ the alphabet is returned to itself and no exchange of statistics occurs.

The same parity is visible directly in the exchange polynomials. Complementing all $d$ descent sets is $\mathcal C_{[d]}$, which by \eqref{eq:majcomp} sends $\maj(T_r)\mapsto M-\maj(T_r)$ with $M=\binom n2$. Each factor contributes one $b^{M}$ and what remains is $b^{-\maj}$ throughout, so
\begin{equation}
 \mathcal C_{[d]}:\quad
 \mathcal P^{+}_{n,d}(\beta)\;\longmapsto\;b^{\,dM}\,\mathcal P^{+}_{n,d}(-\beta),
 \label{eq:coreflip}
\end{equation}
with $b^{-1}=e^{+\beta}$ making the reflection a continuation in the temperature. By Theorem~\ref{thm:dparity} with $|S|=d$ the left side is
$\mathcal P^{-}_{n,d}$ for odd $d$, so
\begin{equation}
 Z^{d}_{n,B}(\beta)=\frac{\mathcal P^{+}_{n,d}(\beta)}{(b)_n^{\,d}},
 \quad
 Z^{d}_{n,F}(\beta)=\frac{b^{\,dM}\,\mathcal P^{+}_{n,d}(-\beta)}{(b)_n^{\,d}},
 \label{eq:ZBZF}
\end{equation}
the two differing only in which end of the same polynomial is read first. Reflecting the denominator as well, $1-b^{-k}=-b^{-k}(1-b^{k})$ gives $(b^{-1})_n=(-1)^n b^{-(M+n)}(b)_n$, so that
\begin{equation}
 Z^{d}_{n,B}(-\beta)=(-1)^{nd}\,b^{\,d(M+n)}\,
 \frac{\mathcal P^{+}_{n,d}(-\beta)}{(b)_n^{\,d}},
\end{equation}
and comparing with \eqref{eq:ZBZF},
\begin{equation}
 Z^{d}_{n,F}(\beta)=(-1)^{n}\,b^{-dn}\,Z^{d}_{n,B}(-\beta),
 \label{eq:oddreflection0}
\end{equation}
only the parity of $n$ surviving since $d$ is odd by assumption. Adding the zero-point energy back, $b^{nd/2}$ at $\beta$ and $b^{-nd/2}$ at $-\beta$, the remaining power cancels and
\begin{equation}
 \;Z^{d}_{n,F}(\beta)=(-1)^{n}\,Z^{d}_{n,B}(-\beta),
 \label{eq:oddreflection}
\end{equation}
for odd values of $d$, which is (11) of Ref.\cite{SchmidtSchnack1999}. This also explains why the authors of Ref.\cite{SchmidtSchnack1999} were able to identify the harmonic oscillator, but no other examples, exhibiting such an odd-dimensional Bose-Fermi temperature transmutation. The relation requires a specialization for which complementing all descent sets is equivalent, up to an overall factor, to $b\mapsto b^{-1}$. Principal specialization has precisely this property, and discovering other such specializations is not trivial.

For even $d$ the same reflection is a symmetry rather than an exchange. Taking $|S|=d$ even in Theorem~\ref{thm:dparity}, the left side of \eqref{eq:coreflip} returns $\mathcal P^{+}_{n,d}$, so
\begin{equation}
 \mathcal P^{\pm}_{n,d}(b)=b^{\,dM}\,\mathcal P^{\pm}_{n,d}(b^{-1}),
 \label{eq:palindromic}
\end{equation}
and both exchange polynomials are palindromic around power of $dM$, i.e. the coefficient of $b^{k}$ equals that of $b^{\,dM-k}$. Reading the list backwards returns the same statistics, which is why no negative temperature relation between $Z_{n,B}$ and $Z_{n,F}$ follows in even dimensions, instead restoring the zero-point gives
\begin{equation}
Z^d_{n,F/B}(\beta) = Z^d_{n,F/B}(-\beta).
\label{eq:evenreflection}
\end{equation}

The degree of $\mathcal P^\pm_{n,d}$ separates the one-dimensional case from the rest. At $d=1$, $\mathcal P^+_{n,1}=1$ and $\mathcal P^-_{n,1}=b^M$, so $Z_{n,F}/Z_{n,B}=b^M$ is a pure power of $b$. The two spectra coincide up to a rigid shift, which is the spectral equivalence of Ref.\cite{Crescimanno}. For $d\ge2$ the exchange polynomials are not monomials and no such shift exists, so the expectation of Ref.\cite{Crescimanno} that spectral equivalence does not persist in higher dimensions is correct even at odd $d$, where \eqref{eq:oddreflection} still relates the two partition functions. The reflection is therefore strictly weaker than spectral equivalence, and the two coincide only at
$d=1$.
 
\subsection{The reflection is not a route around the sign problem}
\label{sec:notasolution}
 
Equation \eqref{eq:oddreflection} does not by itself remove any cancellation.
The reflected result is no longer an ensemble since $e^{+\beta H_1}$ is
not trace-class and the expansion \eqref{eq:psiX} requires $|b|<1$, so
$Z^d_{n,B}(-\beta)$ is a continued rational function. Furthermore, there is no path integral at $-\beta$ since the integrals simply diverge. Evaluating it using the bosonic recursion \eqref{eq:newtonB} at $b\mapsto b^{-1}$ only relocates the sign (since $z_k(b^{-1})=(-1)^{d}b^{kd}z_k(b)$), leaving the precision requirement $B_d\approx\tau(E^d_F-E^d_B)/\ln2$ of Ref.\cite{ChaudharyValenzuelaChin2026} unchanged. For even $d$ the question does not arise at all, since by \eqref{eq:evenreflection} the reflection exchanges nothing. The sum carries the multiplicities $g^{\pm}_d(\boldsymbol\lambda)$, which at $d=3$ are the Kronecker coefficients $\langle\chi^{\lambda_1}\chi^{\lambda_2},\chi^{\lambda_3}\rangle_{S_n}$, and
these are NP-hard to test for nonvanishing \cite{IkenmeyerMulmuleyWalter} and
$\#$P-hard to compute \cite{PakPanova}. The untruncated factorwise sum is
therefore not evaluable term by term in polynomial time, and subtraction-freeness
buys stability without buying speed. But since we still retain the principal specialization for the harmonic case, one might still be able to take advantage of the properties of the spectrum to derive a similar algorithm to B\&Z\cite{BaxterZeilberger2011} for higher dimensions. 

\bibliographystyle{apsrev4-2}
\bibliography{ref}

@article{Ford71,
    author = {Ford, D. I.},
    title = {A Note on the Partition Function for Systems of Independent Particles},
    journal = {Am. J. Phys.},
    volume = {39},
    number = {2},
    pages = {215-220},
    year = {1971},
    doi = {10.1119/1.1986094},
}

@misc{ChaudharyValenzuelaChin2026,
      title={The Path Integral Monte Carlo Sign Problem Is Not Always NP-Hard: Harmonic Fermions Can Be Solved in Quadratic Time}, 
      author={Aarif Chaudhary and Jonas Valenzuela and Siu A. Chin},
      year={2026},
      eprint={2609.09071},
      archivePrefix={arXiv},
      primaryClass={physics.comp-ph},
}

@article{SchmidtSchnack1999,
  author  = {Schmidt, H.-J. and Schnack, J.},
  title   = {Thermodynamic Fermion--Boson Symmetry in Harmonic Oscillator
             Potentials},
  journal = {Physica A},
  volume  = {265},
  pages   = {584--589},
  year    = {1999}
}

@article{SchmidtSchnack2002,
  author  = {Schmidt, H.-J. and Schnack, J.},
  title   = {Partition Functions and Symmetric Polynomials},
  journal = {American Journal of Physics},
  volume  = {70},
  pages   = {53--57},
  year    = {2002}
}

@article{BorrmannFranke1993,
    author = {Borrmann, Peter and Franke, Gert},
    title = {Recursion formulas for quantum statistical partition functions},
    journal = {J. Chem. Phys.},
    volume = {98},
    pages = {2484-2485},
    year = {1993},
    doi = {10.1063/1.464180},
}

@article{Borsens1997,
  title = {Thermodynamics of coupled identical oscillators within the path-integral formalism},
  author = {Brosens, F. and Devreese, J. T. and Lemmens, L. F.},
  journal = {Phys. Rev. E},
  volume = {55},
  pages = {227--236},
  year = {1997},
  doi = {10.1103/PhysRevE.55.227},
}

@article{Knuth1970,
  author  = {Knuth, Donald E.},
  title   = {Permutations, Matrices, and Generalized Young Tableaux},
  journal = {Pacific Journal of Mathematics},
  volume  = {34},
  pages   = {709--727},
  year    = {1970}
}

@book{Macdonald1995,
  author    = {Macdonald, I. G.},
  title     = {Symmetric Functions and Hall Polynomials},
  edition   = {2},
  publisher = {Clarendon Press},
  address   = {Oxford},
  year      = {1995}
}

@book{StanleyEC2,
  author    = {Stanley, Richard P.},
  title     = {Enumerative Combinatorics},
  volume    = {2},
  edition   = {2},
  publisher = {Cambridge University Press},
  address   = {Cambridge},
  year      = {2024}
}

@article{FoataSchutzenberger1978,
  author  = {Foata, Dominique and Sch{\"u}tzenberger, Marcel-Paul},
  title   = {Major Index and Inversion Number of Permutations},
  journal = {Mathematische Nachrichten},
  volume  = {83},
  pages   = {143--159},
  year    = {1978}
}

@misc{BaxterZeilberger2011,
  author        = {Baxter, Andrew and Zeilberger, Doron},
  title         = {The Number of Inversions and the Major Index of
                   Permutations Are Asymptotically Joint-Independently Normal},
  year          = {2011},
  eprint        = {1004.1160},
  archiveprefix = {arXiv},
  primaryclass  = {math.CO}
}

@book{Yau2010,
author = {Yau, Donald},
title = {Lambda-Rings},
publisher = {World Scientific},
year = {2010},
doi = {10.1142/7664},
}

@misc{highestweightvectorstensors,
      title={Highest weight vectors of tensors}, 
      author={Alimzhan Amanov and Damir Yeliussizov},
      year={2026},
      eprint={2504.15413},
      archivePrefix={arXiv},
      primaryClass={math.CO},
}

@misc{newtonsidentity,
      title={Newton's Identity in Finite-Bead Fermionic Partition Function}, 
      author={A. Chaudhary and J. Valenzuela},
      year={2026},
      eprint={2606.05442},
      archivePrefix={arXiv},
      primaryClass={physics.comp-ph},
}

@article{IkenmeyerMulmuleyWalter,
   title={On vanishing of Kronecker coefficients},
   volume={26},
   ISSN={1420-8954},
   DOI={10.1007/s00037-017-0158-y},
   number={4},
   journal={computational complexity},
   publisher={Springer Science and Business Media LLC},
   author={Ikenmeyer, Christian and Mulmuley, Ketan D. and Walter, Michael},
   year={2017},
   month=July, pages={949–992} }

@misc{PakPanova,
      title={On the complexity of computing Kronecker coefficients}, 
      author={Igor Pak and Greta Panova},
      year={2015},
      eprint={1404.0653},
      archivePrefix={arXiv},
      primaryClass={math.CO},
}

@article{LoehrRemmel2011,
  author  = {N. A. Loehr and J. B. Remmel},
  title   = {A computational and combinatorial expos\'e of plethystic calculus},
  journal = {Journal of Algebraic Combinatorics},
  volume  = {33},
  number  = {2},
  pages   = {163--198},
  year    = {2011}
}

@article{Feng_2007,
   title={Counting gauge invariants: the plethystic program},
   volume={2007},
   ISSN={1029-8479},
   DOI={10.1088/1126-6708/2007/03/090},
   number={03},
   journal={Journal of High Energy Physics},
   publisher={Springer Science and Business Media LLC},
   author={Feng, Bo and Hanany, Amihay and He, Yang-Hui},
   year={2007},
   month=Mar, pages={090–090} }

@article{Crescimanno,
  title = {Spectral equivalence of bosons and fermions in one-dimensional harmonic potentials},
  author = {Crescimanno, M. and Landsberg, A. S.},
  journal = {Phys. Rev. A},
  volume = {63},
  issue = {3},
  pages = {035601},
  numpages = {3},
  year = {2001},
  month = {Feb},
  publisher = {American Physical Society},
  doi = {10.1103/PhysRevA.63.035601},
}

@article{Balantekin,
  title = {Partition functions in statistical mechanics, symmetric functions, and group representations},
  author = {Balantekin, A. B.},
  journal = {Phys. Rev. E},
  volume = {64},
  issue = {6},
  pages = {066105},
  numpages = {8},
  year = {2001},
  month = {Nov},
  publisher = {American Physical Society},
  doi = {10.1103/PhysRevE.64.066105},
}

@article{Barghathi2020,
  title = {Theory of noninteracting fermions and bosons in the canonical ensemble},
  author = {Barghathi, Hatem and Yu, Jiangyong and Del Maestro, Adrian},
  journal = {Phys. Rev. Res.},
  volume = {2},
  issue = {4},
  pages = {043206},
  numpages = {16},
  year = {2020},
  month = {Nov},
  publisher = {American Physical Society},
  doi = {10.1103/PhysRevResearch.2.043206},
}

@article{Fock1928,
  author  = {Fock, V.},
  title   = {Bemerkung zur Quantelung des harmonischen Oszillators
             im Magnetfeld},
  journal = {Zeitschrift f\"ur Physik},
  volume  = {47},
  pages   = {446--448},
  year    = {1928},
  doi     = {10.1007/BF01390750}
}

@article{Darwin1931,
  author  = {Darwin, C. G.},
  title   = {The diamagnetism of the free electron},
  journal = {Math. Proc. Cambridge Philos. Soc.},
  volume  = {27},
  pages   = {86--90},
  year    = {1931},
  doi     = {10.1017/S0305004100009373}
}

@article{DRIGHOFILHO2017101,
title = {Superintegrability of the Fock–Darwin system},
journal = {Annals of Physics},
volume = {383},
pages = {101-119},
year = {2017},
issn = {0003-4916},
doi = {https://doi.org/10.1016/j.aop.2017.05.003},
author = {E. Drigho-Filho and S. Kuru and J. Negro and L.M. Nieto},
}

@misc{ChinChadPara,
      title={Parafermions in plain sight}, 
      author={Siu A. Chin and A. Chaudhary},
      year={2026},
      eprint={2609.00545},
      archivePrefix={arXiv},
      primaryClass={cond-mat.stat-mech},
}

@misc{Wang2026,
      title={On $R$-parastatistics I: Foundation}, 
      author={Zhiyuan Wang and Kaden R. A. Hazzard},
      year={2026},
      eprint={2607.26351},
      archivePrefix={arXiv},
      primaryClass={quant-ph},
}

@article{Wang2025,
   title={Particle exchange statistics beyond fermions and bosons},
   volume={637},
   ISSN={1476-4687},
   DOI={10.1038/s41586-024-08262-7},
   number={8045},
   journal={Nature},
   publisher={Springer Science and Business Media LLC},
   author={Wang, Zhiyuan and Hazzard, Kaden R. A.},
   year={2025},
   month=Jan, pages={314–318} 
}

\end{document}